\documentclass[aps,showpacs]{revtex4}

\usepackage{subfig}
\usepackage{color}
\usepackage{xcolor,eso-pic,graphicx}
\usepackage{pifont} 
\usepackage{amsmath}
\usepackage{empheq}
\usepackage{fancybox}
\usepackage{amssymb}
\usepackage{graphics}
\usepackage{epsfig}
\usepackage{caption}
\usepackage{float}
\usepackage{alltt}
\begin{document}
\renewcommand{\floatpagefraction}{7.0}
\begin{flushright}
March 27, 2014 
\par
\end{flushright}

\title{A Singularity-Free Coordinate System for X-point Geometries}

\author{F. Hariri$^{1,*}$, P. Hill$^{1}$, M. Ottaviani$^{1}$, Y. Sarazin$^{1}$}

\affiliation{$^{1}$CEA, IRFM, F-13108 Saint-Paul-lez-Durance, France}

%%****************************************
% Abstract
%%****************************************

\begin{abstract}
A Flux Coordinate Independent (FCI) approach for anisotropic systems, not based on magnetic flux coordinates has been introduced in [F. Hariri and M. Ottaviani, Comput. Phys. Commun., {\bf 184}, 2419 (2013)]. In this paper, we show that the approach can tackle magnetic configurations including X-points. Using the code FENICIA, an equilibrium with a magnetic island has been used to show the robustness of the FCI approach to cases in which a magnetic separatrix is present in the system, either by design or as a consequence of instabilities. Numerical results are in good agreement with the analytic solutions of the sound-wave propagation problem. Conservation properties are verified. Finally, the critical gain of the FCI approach in situations including the magnetic separatrix with an X-point is demonstrated by a fast convergence of the code with the numerical resolution in the direction of symmetry. The results highlighted in this paper show that the FCI approach should be able to address turbulent transport problems in X-point geometries. 
\end{abstract}

\maketitle

% Present Address
\begin{flushleft}
\footnotesize{\textit{$^*$E-mail address:} Farah.Hariri@epfl.ch\\\textit{Present address: EPFL - SB - CRPP-TH Station 13, CH-1015 Lausanne, Switzerland}}
\end{flushleft}

%%****************************************
%% Define the path for the Pictures Folder
\def\RepFigures{FIGURES_EPS}
%%****************************************

%%%%%%%%%%%%%%%%%%%%%%%%%%%%%%%%%%%%%%%%%%%%%%
\section{Introduction}
\label{sec:Introduction}
%%%%%%%%%%%%%%%%%%%%%%%%%%%%%%%%%%%%%%%%%%%%%%
The presence of a magnetic field in plasmas is known to introduce a strong anisotropy in the system. This is is commonly met in fusion and most astrophysical plasmas. Developing numerical codes that take into account the strong plasma anisotropy and make efficient use of computational resources is crucial, for instance, to simulate ITER-like tokamak plasmas. Efforts have been made in this direction that lead to introducing optimized coordinate systems, so-called field-aligned coordinates, that take advantage of this anisotropy. Two-dimensional field-aligned coordinates, based on magnetic flux variables, were first presented in~\cite{Cowley19912767,Hammett1993973,Beer19952687,Dimits19934070, Scott19982334, Scott2001447}. They have a fundamental drawback of not handling situations including the magnetic separatrix, due to the singularity of the field-aligned metric. The approach proposed later in~\cite{Ottaviani20111677} avoids this shortcoming, but still relies on flux coordinates, for instance ($r,\theta$). A generalized 3-dimensional coordinate system, referred to as the Flux Coordinate Independent (FCI) system, not based on magnetic flux coordinates, is introduced in~\cite{Hariri20132419} and elaborated in~\cite{Hariri2013FENICIA}. The FCI coordinate system is constructed in a way that avoids the use of flux variables to discretize the fields in the poloidal plane. As discussed in~\cite{Hariri20132419, Hariri2013FENICIA},  it allows the parallel derivative to be constructed by tracing the magnetic field lines from one perpendicular plane to the next, and interpolating to find the desired quantity. This frees us from using flux coordinates in the perpendicular plane, thus allowing for complex magnetic geometries free of singularities.  The aim of the present work is to highlight the advantage underlying the use of the FCI field-aligned system for X-point magnetic configurations. Examples of such configurations are magnetic islands (driven by magnetohydrodynamics modes or embedded in the magnetic equilibrium such as in stellerators) and axi-symmetric X-points on the last closed field surface (LCFS) of tokamak plasmas.\\

These configurations, found in fusion as well as astrophysical plasmas, have a singularity in the metric of a coordinate system that uses the magnetic flux as a coordinate. This difficulty originates in the nature of the flux function, which has a saddle point (known as X-point) in one or more locations. Using the flux coordinate independent (FCI) approach avoids this problem and opens the way for the numerical simulations of plasma turbulence in a tokamak geometry encompassing both closed and open magnetic field line regions. Firstly, this is a necessary feature of codes addressing the question of the transition from the low (L) to the high (H) confinement regime in X-point geometry. It is worth reminding that ITER would operate in the H confinement mode to achieve the expected thermonuclear fusion power. Understandably, research in this direction is a hot topic and understanding turbulence in X-point geometries is crucial for predicting the L-H power threshold. Secondly, studies of the interaction between micro-turbulence and magnetic islands exhibit a significant interplay affecting their dynamics. The most recent investigations on this topic are reported in~\cite{Militello2008Interaction, Hornsby2012On, ishizawa2013magnetic}.\\ 

The ultimate result of this paper is showing that the FCI approach allows, in particular, not only a more natural treatment of the operations in the poloidal plane, but it also easily deals with X-point configurations and with O-points such as the magnetic axis, since it is constructed on coordinate systems with non-singular metric. For this purpose, we use the FENICIA code presented in~\cite{Hariri20132419, Hariri2013FENICIA}, where the {\bf F}lux indep{\bf EN}dent f{\bf I}eld-aligned {\bf C}oord{\bf I}nate {\bf A}pproach has been implemented. FENICIA is based on a modular numerical scheme specially designed to address the anisotropic transport of any set of equations belonging to the general class of models $(1)$ in~\cite{Hariri20132419, Hariri2013FENICIA}.  Since this code is easily extensible to different meshes and coordinate systems,  we use it here to demonstrate the application of the FCI coordinate system to a magnetic island in a slab version of the code. The outline of this paper is as follows. First of all, analytic solutions of a sound-wave propagation model for an island magnetic equilibrium are presented in Section~\ref{sec:interior_exterior_solutions}. Then numerical tests are performed at the exterior and at the interior of the island to show the convergence of the numerical solution to the analytic one in Section~\ref{sec:numerical_tests}. Across the separatrix, where there are not yet known analytic solutions, two conclusive numerical convergence tests are presented in Section~\ref{sec:tests_separatrix} allowing us to show the robustness of the FCI approach in X-point and O-point magnetic configurations.

%***********************************************************************
\section{Sound wave model in a magnetic island geometry}
\label{sec:interior_exterior_solutions}
%***********************************************************************
Consider a class of static low-$\beta$ equilibria, such that the suitably normalized axisymmetric magnetic field is given by
\begin{equation}
\mathbf{B} = \mathbf{b(x)} + \mathbf{\hat{z}}
\label{mag-field}
\end{equation}
where one employs a three dimensional Cartesian reference system $(x,y,z)$ such that $\mathbf{\hat{z}}$ is the direction of the magnetic axis, the main magnetic field along $z$ is constant and normalized to unity, and $\mathbf{b(x)}$ is the poloidal magnetic field in the poloidal plane $(x,y)$. The two-dimensional vector $\mathbf{x}$ indicates the position in this plane. The poloidal field can be written in terms of a flux function $\psi(\mathbf{x})$ such that
\begin{equation}
\mathbf{b} = \pmb\nabla \times (\psi \mathbf{\hat{z}})   
\label{pol-field}
\end{equation}     
Magnetic surfaces can be labeled by the value of $\psi$. Both closed and open field lines can be treated. As extensively discussed in~\cite{biskamp1997nonlinear}, slowly growing \emph{tearing} instabilities reconnect magnetic flux surfaces to form magnetic islands. Analytical methods were derived to examine the linear stability and radial distribution of tearing modes~\cite{furth2003tearing, cowley1988electron, Cowley1985Some, connor1990micro}. The nonlinear growth of these modes was also investigated in~\cite{rutherford1973nonlinear} and extended to cases where current nonlinearities are important as shown in~\cite{Waelbroeck1993494}.\\

Consider a magnetic equilibrium characterized by a magnetic island whose half radial width is equal to $\delta_x = 2\sqrt{A}$, with $A$ being a parameter. Such an equilibrium corresponds to a magnetic field given by Eqs.~\eqref{mag-field}-\eqref{pol-field}, with:
\begin{equation}
  \psi(x,y) = - \frac{(x-1)^2}{2} + A \cos(y)
\end{equation}
The considered domain, for an island centered at $x=1$, is $x_{min} \leq x \leq x_{max}$ (in practice, $x_{max}=1+\Delta$ and $x_{min}=1-\Delta$, with $0<\Delta<1$) and $-\pi \leq y \leq \pi$, with $0\leq z\leq 2\pi$. \\

Let us consider the following sound-wave model that propagate parallel to the magnetic field
\begin{eqnarray}\label{DW2_slab}
  && \partial_t \phi + C \, \nabla_\parallel u = 0 \\\nonumber
  && \partial_t u +  \frac{C(1+\tau)}{\tau}\; \nabla_\parallel \phi = 0 
\end{eqnarray}
$\phi$ is the electrostatic potential and $u$ is the wave's parallel velocity. We define two dimensionless parameters: $C = a/R \times 1/\rho_*$ where $a$ is the tokamak minor radius, $R$ is the tokamak major radius, $\rho_* = \rho_s / a$ is the reduced gyro-radius with $\rho_s$ being the ion sound Larmor radius; and $\tau=1$. The parallel gradient operator is defined as follows:
\begin{equation}
  \nabla_\parallel = - [\psi, \cdot ] + \partial_z
\end{equation}
and the Poisson bracket is defined by $[\psi, \cdot ] = \partial_x \psi \partial_y - \partial_y \psi \partial x$.
The aim of the present section is to construct analytic solutions of model~\eqref{DW2_slab} in the magnetic equilibrium defined above, i.e. in the presence of a magnetic island. 
To do so, it is convenient to employ flux coordinates. It readily appears that $\psi$ is a label of magnetic surfaces, since $\nabla_\parallel \psi=0$. \\

Let us introduce the following new set of coordinates:
\begin{eqnarray}
  \label{eq:rho_eta}
  \rho &=& - \psi/A = \frac{(x-1)^2}{2A} - \cos y\\
  \eta &=& g(\rho) \int_0^y \frac{dy'}{[\cos y' + \rho]^{1/2}}  
\end{eqnarray} 
The coordinate $\eta$ is a straight-field-line angle-like variable and has still to be calculated. As detailed in~\ref{app_A}, for $\eta$ to be an angle ranging from $-\pi$ to $\pi$, $g(\rho)$ has to be given different expressions depending on whether $\rho$ is bigger or smaller than 1, $\rho=1$ being the radius at the separatrix of the island. There are indeed three regions, each with its own specific expression of $g$ and $\eta$. The interior region of the island corresponds to values of $\rho <1$, while $\rho>1$ characterizes the exterior region of the island. One can actually distinguish two exterior regions: the left side and the right side of the island. $g(\rho)$ in both of the exterior regions has the following expression:
\begin{equation}
  g(\rho)|_{\rho>1} = \pm \frac{\pi}{2} (1+\rho)^{1/2}\;
    \left[ K\left(\frac{2}{1+\rho}\right) \right]^{-1}
\end{equation}
where the sign $\pm$ depends on whether one considers the right side (+) or the left side (-) of the exterior region (cf.~\ref{app_A}). $K$ is the elliptic integral of the first kind:
\begin{equation}
  K(x) \equiv \int_0^{\pi/2} d\theta\; (1-x\sin^2\theta)^{-1/2}
\end{equation}
Conversely, $g(\rho)$ can be shown to be as follows for the interior region $\rho<1$:
\begin{equation}\label{g_rho}
  g(\rho)|_{\rho<1} = \frac{\pi}{2\sqrt{2}} \;
    \left[ K\left(\frac{1+\rho}{2}\right) \right]^{-1}
\end{equation}
To illustrate this, contour plots of $\rho$ and $\eta$ are shown in figures~\ref{rho}-\ref{eta} where $\rho$ defines the island geometry and $\eta$ defines the inner and outer regions of the island. 
%\begin{figure}[h!]	
%  \centering
%  \subfloat[]{\includegraphics[width=8cm, height=7cm]{\RepFigures/FIG1.eps}\label{rho}}\hspace*{1mm}
%  \subfloat[]{\includegraphics[width=8cm, height=7cm]{\RepFigures/FIG2.eps}\label{eta}} 
%  \caption{The new coordinates $\rho$ (a) and $\eta$ (b) as a function of the grid mesh coordinates $(x,y)$.}
%\end{figure}
It can be shown that the parallel gradient takes the following expression in terms of this new set of coordinates:
\begin{equation}
  \nabla_\parallel = g(\rho)\sqrt{2A}\,\frac{\partial}{\partial \eta} 
    + \frac{\partial}{\partial z}
\end{equation}
Let us then look for wave-like solutions of model~\eqref{DW2_slab} of the form:
\begin{eqnarray} \label{eq:X_exact_solutions}
  \left(\begin{array}{c}
  \phi(\rho,\eta,t) \\ u(\rho,\eta,t) \end{array} \right)
  =
  \left(\begin{array}{c}
  \phi_0(\rho) \\ u_0(\rho) \end{array} \right)
  \cos\left[m\eta-nz-\omega(\rho)t\right]
\end{eqnarray}
with $(m,n)$ standing for the wave numbers in $\eta$ and $z$ respectively, and $\omega(\rho)$ being the mode frequency. Injecting these expressions in Eq.~\eqref{DW2_slab} leads to the following system:
\begin{eqnarray}
  \left(\begin{array}{cc}
  -\omega & C \left[ g(\rho)\sqrt{2A} m -n \right]\\ 
  \frac{C(1+\tau)}{\tau}\; \left[ g(\rho)\sqrt{2A} m -n \right] & -\omega \end{array} \right)
  \left(\begin{array}{c}
  \phi_0(\rho) \\ u_0(\rho) \end{array} \right) = 0
\end{eqnarray}
The eigenfrequency solution of the dispersion relation depends on $\rho$:
\begin{equation} \label{eq:omegapm}
  \omega^\pm_{mn} (\rho)= \pm C\left(\frac{1+\tau}{\tau}\right)^{1/2}\; 
  \left[ g(\rho)\sqrt{2A} m -n \right]
\end{equation}
The eigenvectors are then 
\begin{equation} \label{eq:u0}
  u_0(\rho) = \frac{C(1+\tau)}{\tau \omega^\pm_{mn} (\rho)}\; \phi_0(\rho) 
\end{equation}
For a given expression of $\phi_0(\rho)$, $u_0$ can be calculated by using Eq.~\eqref{eq:u0}, with $\omega^\pm_{mn} (\rho)$ given by Eq.~\eqref{eq:omegapm}. Then, Eq.~\eqref{eq:X_exact_solutions} provides analytic solutions of the model~\eqref{DW2_slab}, which will be compared to their numerical counterparts in the next section. While the above solutions can be derived for $\rho<1$ and $\rho>1$, there is no obvious analytic solution to this problem at $\rho=1$. This issue is left for future investigation.
%***********************************************************************
\section{Numerical tests at the exterior and interior of the island}
\label{sec:numerical_tests}
%***********************************************************************
The first test aims at verifying that the numerical simulation of sound-wave propagation outside the X-point gives results in good agreement with the analytic solutions of this problem up to the order of accuracy of the chosen numerical scheme. FENICIA is used with a version embedding an island magnetic equilibrium. With respect to the model presented in~\cite{Hariri20132419, Hariri2013FENICIA}, Dirichlet boundary conditions are still used in the $\mathit{x}$ direction, but periodic boundary conditions are now used in the $\mathit{y}$ and $\mathit{z}$ directions. The results of our investigation are presented into the influence of a static axi-symmetric magnetic island, including the X- and O- points, on sound-waves. We consider a box of size $2 \Delta = 0.2$ (normalized to the radial position of the separatrix), an island of size $4\sqrt{A}$ with $A = 10^{-3}$ and a mode $(m,n)$ resonant at $\rho = \rho_{mn}$, i.e: $m/n =  1/(g(\rho_{mn}) \, \sqrt{2\,A})$. This means that $k_\parallel = 0$ at $\rho_{mn}$. For this value of $A$, we show $1/(g(\rho_{mn}) \, \sqrt{2\,A})$ as a function of $\rho$ for both the exterior and the interior of the island as shown in Fig.~\ref{mn_vs_rho}. For rational values of $1/(g(\rho) \, \sqrt{2\,A})$, there exists a resonant $(m,n)$. One sees that the position around $\rho =1$ is directly constrained by high $m$ or high $m/n$ values, especially if one wishes to get closer to the separatrix starting from the interior of the island.
%\begin{figure}[h!]	
%  \centering
%  \includegraphics[width=7cm, height=6cm]{\RepFigures/FIG3.eps}
%  \caption{plot of resonant $m/n$ as a function of $\rho$ for both the exterior and the interior of the island}
%  \label{mn_vs_rho}
%\end{figure}
We initialize a perturbation with the form
\begin{equation}
\phi(t=0) = \phi_0(\rho ) \cos( m\eta - nz)
\end{equation}
where $\phi_0(\rho)$ is a Gaussian structure centered around $\rho_{mn}$ and having the following form
\begin{equation}
\phi_0 (\rho) = \exp{ \left [{- \frac{(\rho-\rho_{mn})^2}{\Delta \rho^2}} \right ]} \times \left (\frac{\rho}{\rho_{mn}} \right )^m \times \left (\frac{\rho-\rho_{bd}}{\rho_{mn}-\rho_{bd}}\right )^2
\end{equation}
with $\Delta \rho = \rho_{mn}/m$ and $\rho_{bd}$ is the value of $\rho$ at the boundaries $x = \{x_{min},x_{max} \}$ for $y=0$. The small radial width of the envelope $\phi_0$, around the resonant surface $\rho_{mn}$, ensures that $k_\parallel \ll 1$. Two cases are to be examined: the exterior of the island $(\rho > 1)$, and the interior $(\rho < 1)$. The only difference between the two cases comes from the two different expressions of $g(\rho)$ given by Eq.~\eqref{g_rho} in the previous section. For the former, we consider a perturbation $\phi_0(\rho)$ centered around $\rho_{mn} = 1.25$ for $(m,n) = (24,1)$. The initial condition for a simulation of size $(n_x, n_y, n_z) = (800 \times 800 \times 20)$ is shown in Fig.~\ref{init_exterior}. Note that time $t$ is normalized to the Bohm timescale (which is long by a factor $1/\rho_*$ compared to the significant  turbulence time) and the time step considered for the following simulations is $\Delta t = 10^{-3}$. In Fig~\ref{final_exterior}, we show the solution at time $t=1$ . For clarity, a zoomed view of the initial condition and the final solution are also shown in Figs.~\ref{init_exterior_zoom}-\ref{final_exterior_zoom}.
%\begin{figure}[h!]	
%  \centering
%  \subfloat[]{\includegraphics[width=5.5cm, height=5cm]{\RepFigures/FIG5.eps}\label{init_exterior}}
%  \subfloat[]{\includegraphics[width=5.5cm, height=5cm]{\RepFigures/FIG6.eps}\label{final_exterior}} \\
%  \subfloat[]{\includegraphics[width=5.5cm, height=5cm]{\RepFigures/FIG7.eps}\label{init_exterior_zoom}}
%  \subfloat[]{\includegraphics[width=5.5cm, height=5cm]{\RepFigures/FIG8.eps}\label{final_exterior_zoom}} 
%  \caption{a) Initial condition $\phi(t=0)$; (b) solution $\phi$ at the final time step $t=1$; (c) zoomed view of the initial condition $\phi(t=0)$; (d) zoomed view of the solution $\phi$ at the final time step $t=1$}
%\end{figure}
The emphasis is now on showing that the exact slab solution given by~\eqref{eq:X_exact_solutions} is recovered at the exterior of the island where $\rho > 1$. In Fig.~\ref{errors_exterior}, we plot the relative error 
\begin{equation}
\frac{\left< (\phi_{num} - \phi_{exact})^2 \right>^{1/2}}{\left< (\phi_{exact})^2 \right>^{1/2}}
\end{equation} 
between the exact and the numerical solution as a function of time for different spatial resolutions $(n_x,n_y,n_z) = \{(400, 400,20); (600, 600,20); (800, 800,20) \}$ where $\left< \cdot \right> = \int_0^{2\pi} dy \, \int_0^{2\pi} dz \, \int_1^{x_{max}} dx$. The volume integral is defined over half of the domain (right side of the island). The calculations give similar results for the left side of the island. From Fig.~\ref{errors_exterior}, we see that the numerical results converge quickly to the analytic solution. 
%\begin{figure}[h!]	
%  \centering
%  \includegraphics[width=6cm, height = 6cm]{\RepFigures/FIG9.eps}
%  \caption{errors as a function of time given by different spatial resolutions}
%  \label{errors_exterior}
%\end{figure}
The same tests are performed at the interior of the island where we consider a perturbation $\phi_0(\rho)$ centered around $\rho_{mn} = 0.58$ for $(m,n) = (45,1)$. The initial condition for a simulation of size $(800 \times 800 \times 20)$ is shown in Fig.~\ref{init_interior} and the final solution at $t=1$ is given in Fig~\ref{final_interior}. A zoomed view of the initial condition and the final solution are also shown in Figs.~\ref{init_interior_zoom}-\ref{final_interior_zoom}.
%\begin{figure}[h!]	
%  \centering
%  \subfloat[]{\includegraphics[width=5.5cm, height=5cm]{\RepFigures/FIG10.eps}\label{init_interior}}
%  \subfloat[]{\includegraphics[width=5.5cm, height=5cm]{\RepFigures/FIG11.eps}\label{final_interior}}\\
%  \subfloat[]{\includegraphics[width=5.5cm, height=5cm]{\RepFigures/FIG12.eps}\label{init_interior_zoom}}
%  \subfloat[]{\includegraphics[width=5.5cm, height=5cm]{\RepFigures/FIG13.eps}\label{final_interior_zoom}} 
%  \caption{(a) Initial condition $\phi(t=0)$; (b) solution $\phi$ at the final time step $t=1$; (c) zoomed view of the initial condition $\phi(t=0)$; (d) zoomed view of the solution $\phi$ at the final time step $t=1$}
%\end{figure}
The numerical solution is again compared to the analytic solution given by~\eqref{eq:X_exact_solutions} for $\rho < 1$ . This is illustrated in Fig.~\ref{errors_interior} where we plot the relative error as a function of time. We observe that the numerical solution to the sound-wave problem converges to the one derived analytically as the grid size $\Delta x$ decreases. However, the relative error for the resolutions considered here is still high ($\sim 30\%$). We interpret this to indicate that the poloidal resolution is not sufficient to be able to resolve the $m=45$ mode at the interior of the island. 
%\begin{figure}[h!]	
%  \centering
%  \includegraphics[width=6cm, height = 6cm]{\RepFigures/FIG14.eps}
%  \caption{errors as a function of time given by different spatial resolutions}
%  \label{errors_interior}
%\end{figure}
%***********************************************************************
\section{Tests across the separatrix}
\label{sec:tests_separatrix}	
\begin{figure}[h!]	
  \centering
 \hspace*{-0.5cm}  \vspace*{-5mm}\subfloat[t=0]{\includegraphics[width=6cm, height = 5.3cm]{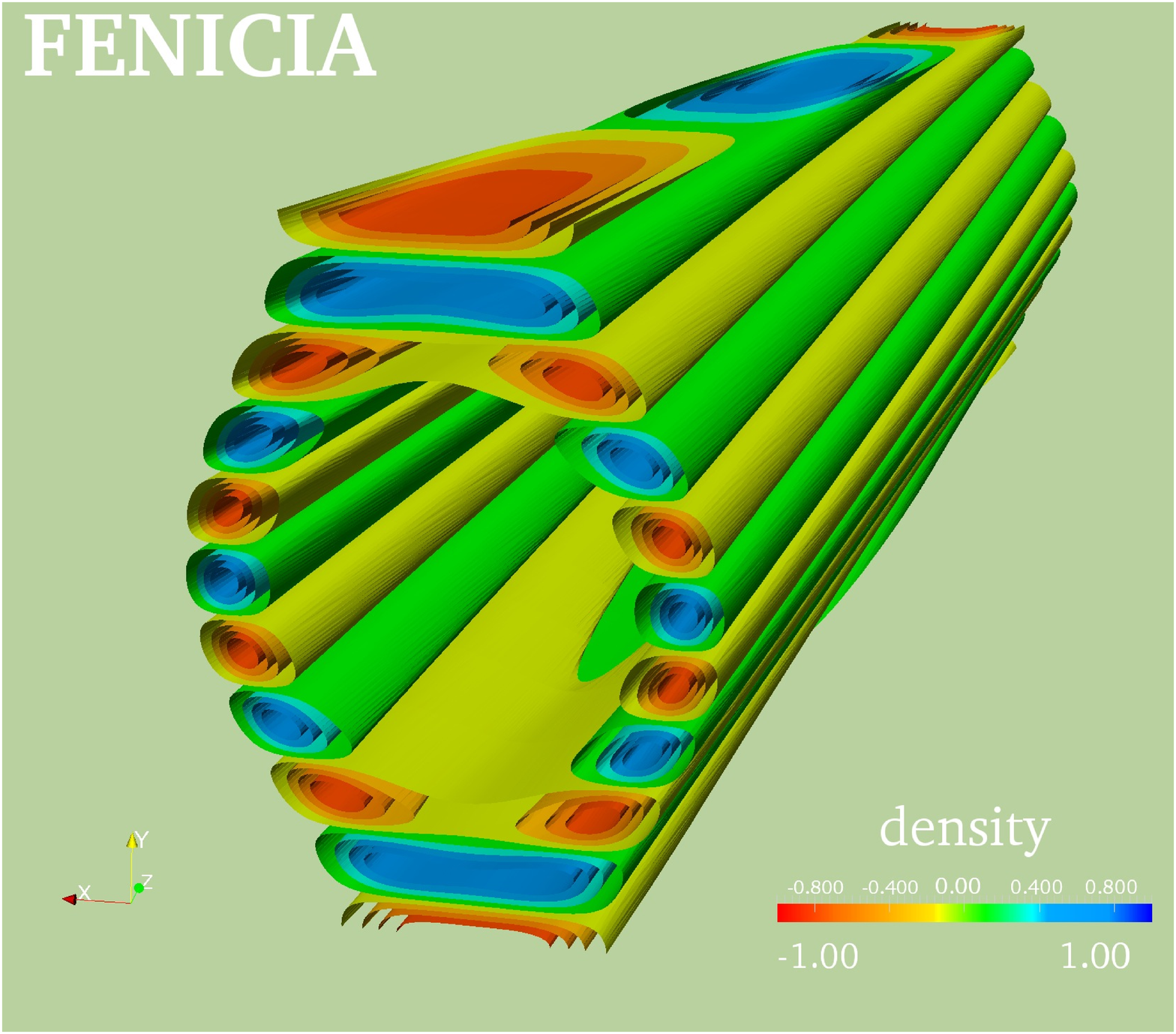}} \hspace*{3ex}
  \hspace*{-0.5cm} \subfloat[t=1]{\includegraphics[width=6cm, height = 5.3cm]{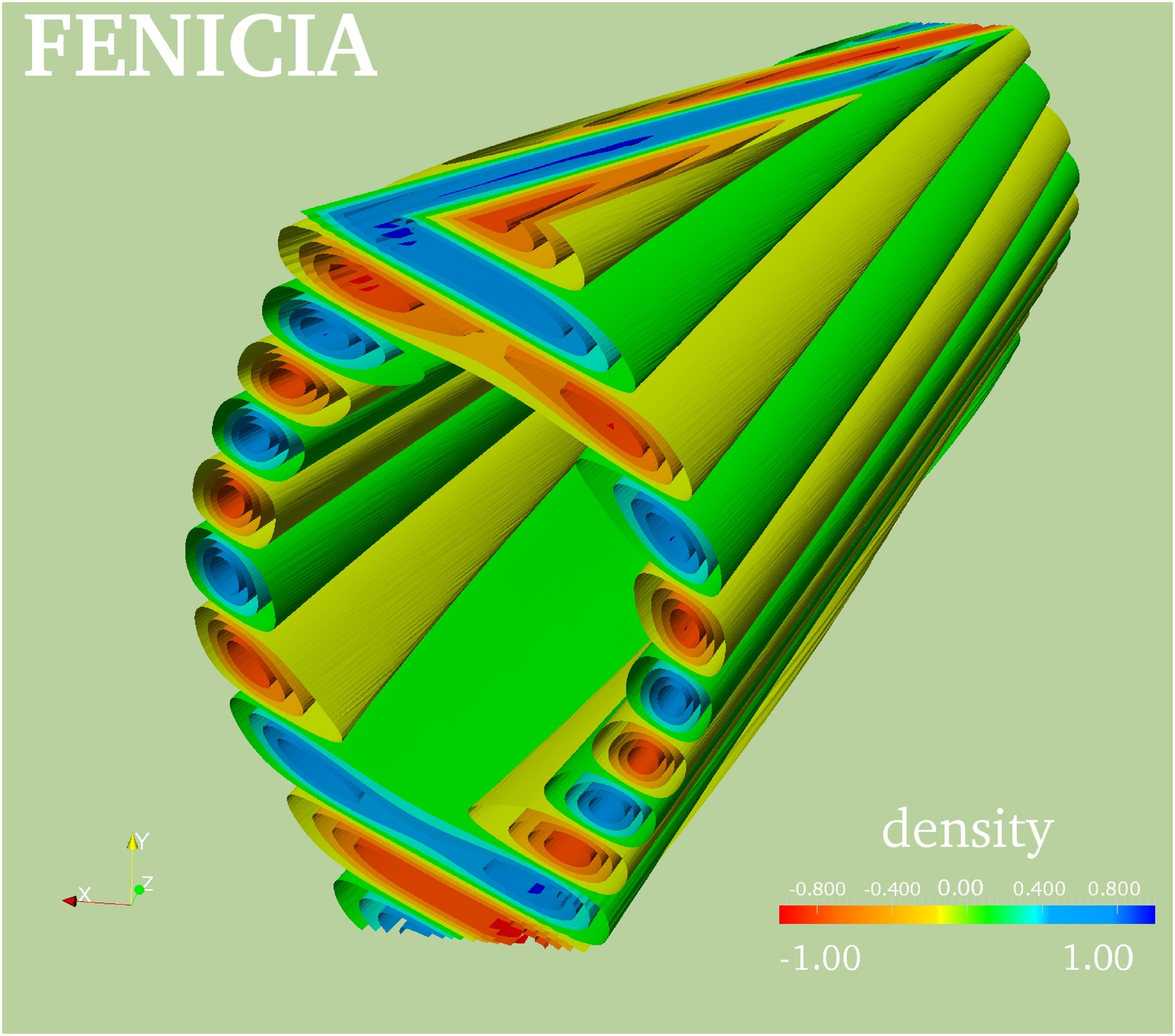}}\\ \vspace*{1ex}
  \hspace*{-0.5cm} \subfloat[t=2]{\includegraphics[width=6cm, height = 5.3cm]{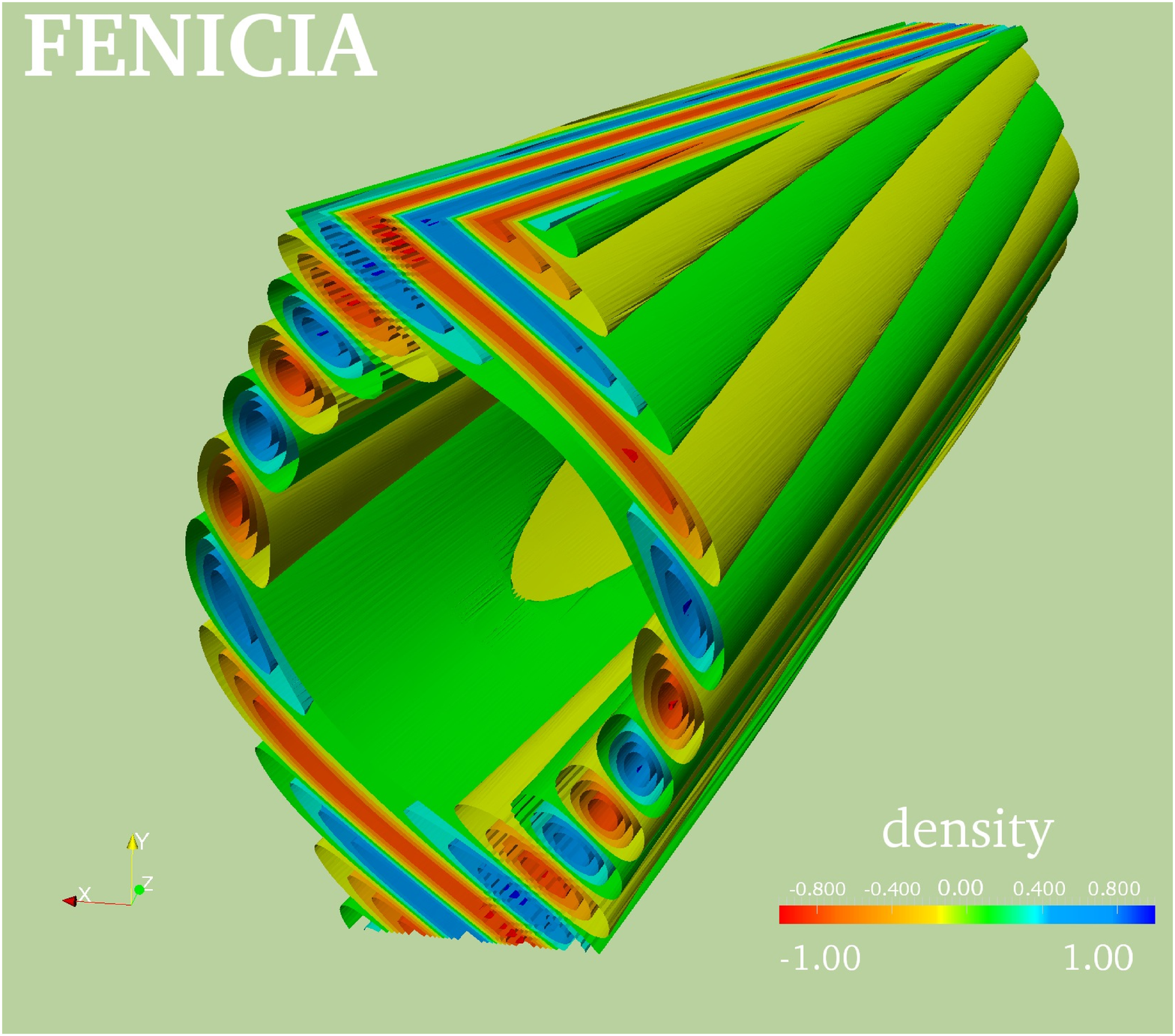}} \hspace*{3ex}
  \hspace*{-0.5cm} \subfloat[t=3]{\includegraphics[width=6cm, height = 5.3cm]{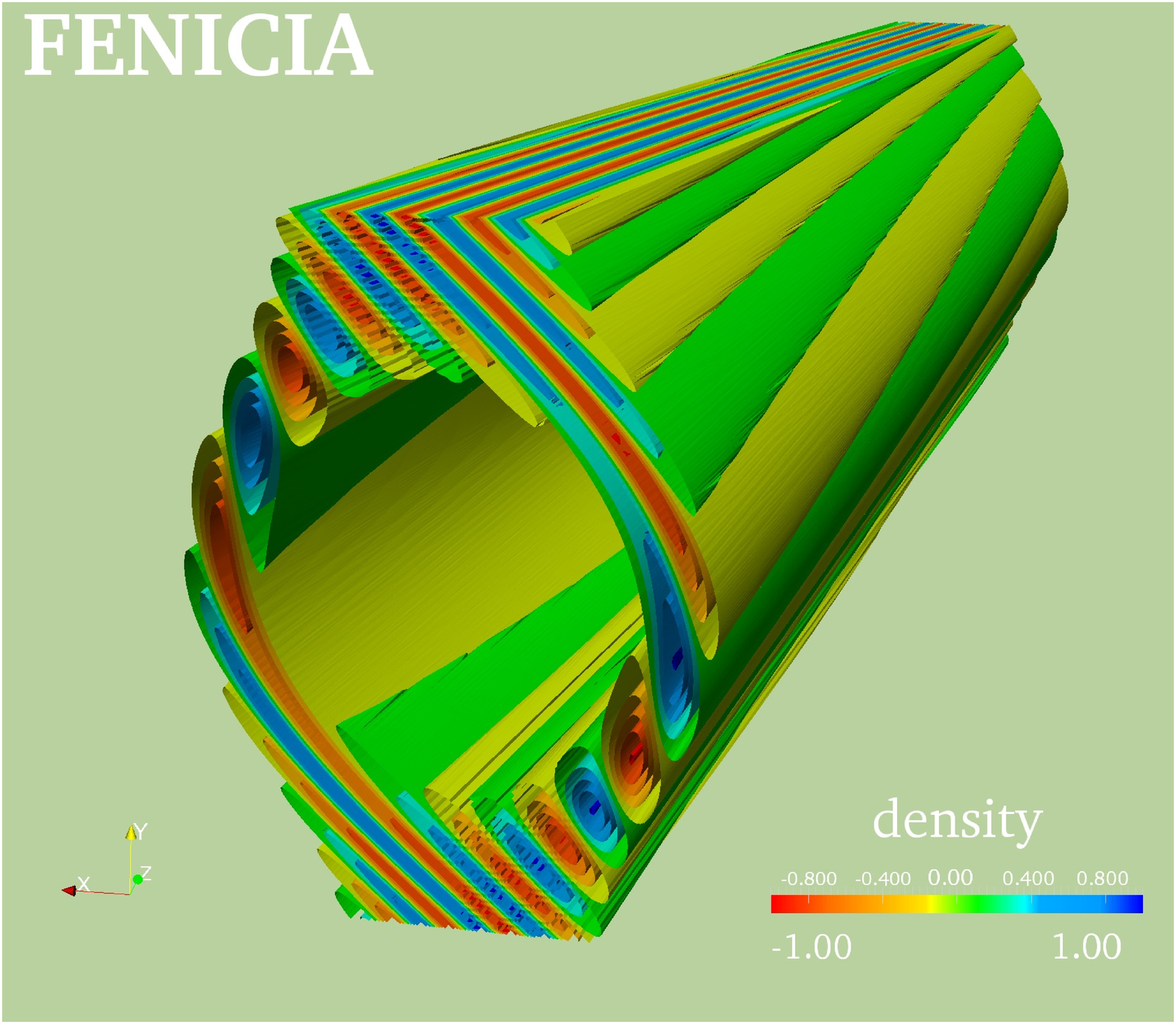}}\\
  \hspace*{-0.5cm} \subfloat[t=4]{\includegraphics[width=6cm, height = 5.3cm]{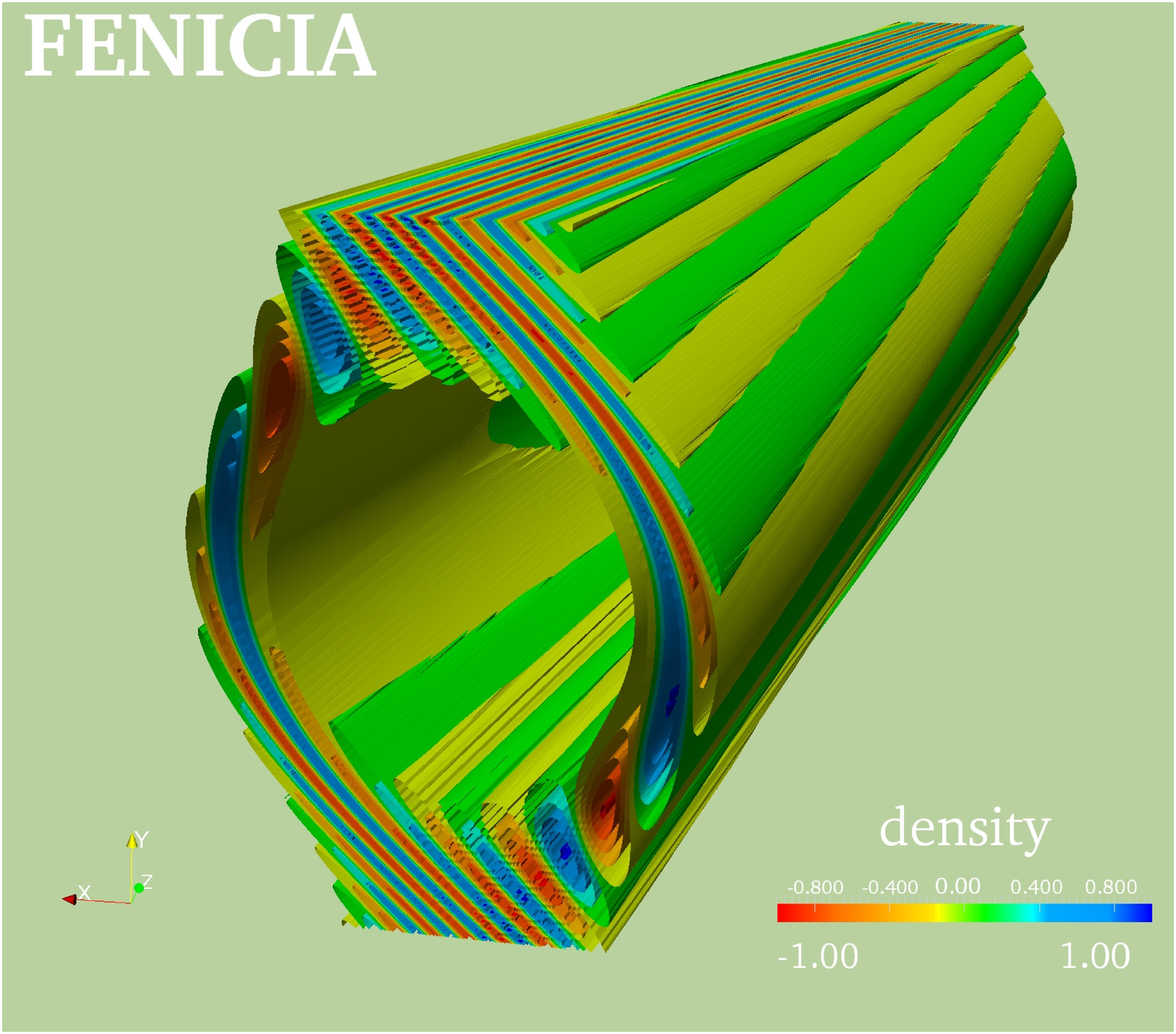}} \hspace*{3ex}
  \hspace*{-0.5cm} \subfloat[t=5]{\includegraphics[width=6cm, height = 5.3cm]{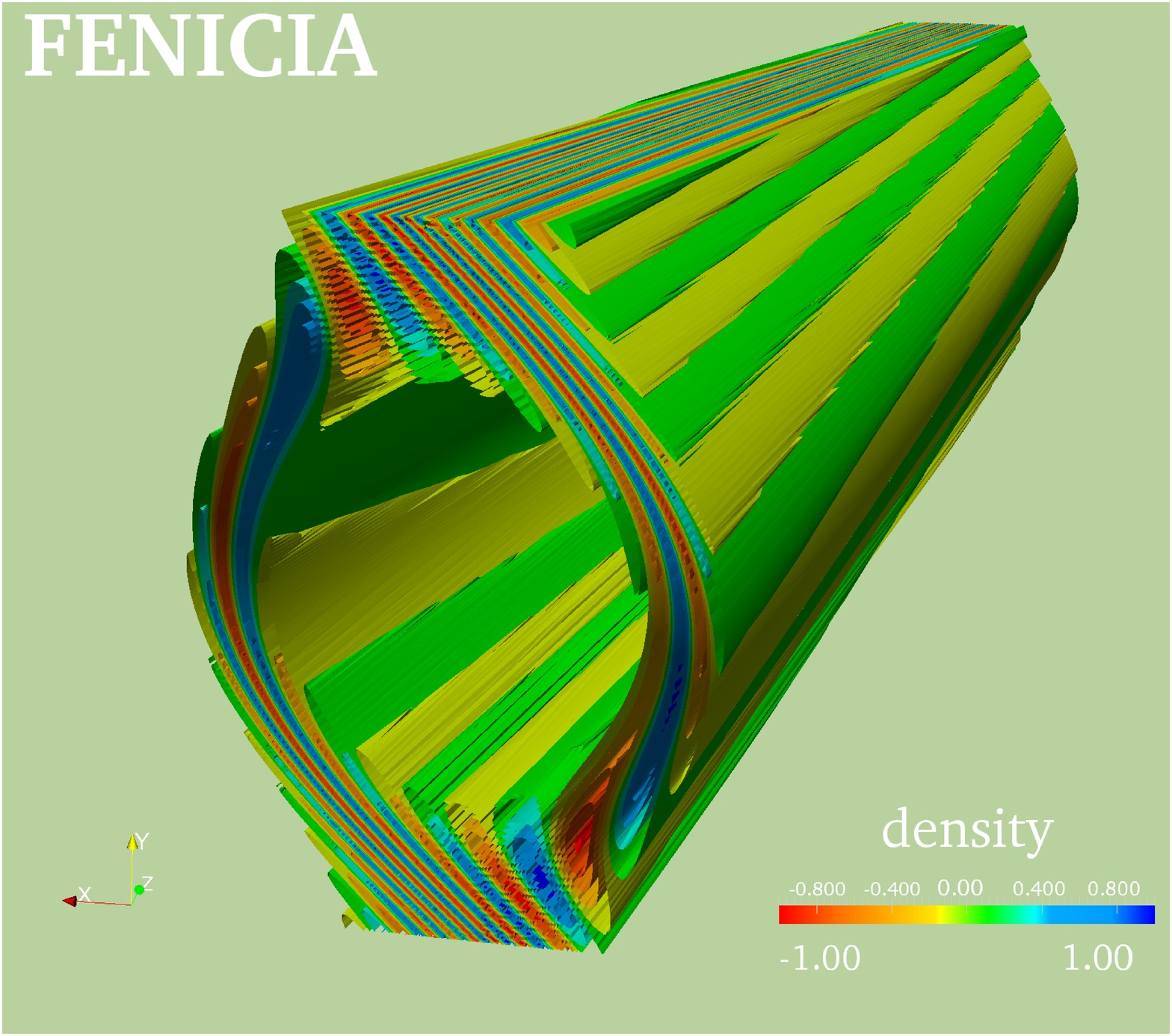}}\\
  \caption{$3$D snapshots of potential fluctuations at different simulation times}
  \label{3D_snapshots_slab}
\end{figure}

%***********************************************************************
At the separatrix ($\rho = 1$), the analytic solutions~\eqref{eq:X_exact_solutions} are no longer valid. As stated in section~\ref{sec:interior_exterior_solutions}, we do not have an obvious analytic solution at the separatrix. Given this limitation, rather than considering an initial condition as a function of the variable $\eta$, which diverges at $\rho = 1$, we instead consider a perturbation crossing the separatrix of the form
\begin{equation}
\phi(t) = N_0(\rho ) \cos( my - nz)
\end{equation}
where $N_0(\rho)$ is a Gaussian structure given by
\begin{equation}
N_0 (\rho) = \mathrm{e}^{-(\rho - 1)^2/\Delta \rho^2}
\end{equation}
where $\Delta \rho$ is a parameter set to $\Delta \rho = 0.5$ here for the perturbation to cover both the left and right-hand sides of the separatrix. A series of simulations were then run with $\Delta = 0.1$, $A = 10^{-3}$ and $(m,n) = (5,1)$. Figures~(\ref{2D_snapshots_potential_slab}-\ref{3D_snapshots_slab}) show the evolution of the electrostatic potential as a function of time (normalized to Bohm time). The time step for this simulation is $\Delta t = 10^{-3}$ and the box size is $(200 \times 200 \times 20)$. From Fig.~\ref{2D_snapshots_potential_slab} one sees that modes across the separatrix gradually evolve and shear. Due to the periodicity of the simulation box, we also observe eddies leaving one side of the box and reentering on the other side of the box. $3$D illustrations of the same simulation are shown in Fig.~\ref{3D_snapshots_slab} allowing us to see the elongated nature of structures along the magnetic field lines. It is also evident from that figure that the direction parallel to the magnetic field does not coincide with the direction of symmetry ($\mathit{z}$).  
%\begin{figure}[h!]	
%  \centering
%  \includegraphics[width=\textwidth ]{\RepFigures/FIG15.eps}
%  \caption{$2$D snapshots of potential fluctuations at different simulation times}
%  \label{2D_snapshots_potential_slab}
%\end{figure}
Because our ultimate goal was to prove the validity of the FCI approach for X-point geometries, three main tests are to be considered hereafter. The first test consists in showing the conservation of an energy-like quadratic quantity of the sound-wave model~\eqref{DW2_slab}. For this purpose, we perform a scan in $n_x$ and in $n_z$, the number of points in the $\mathit{x}$ and $\mathit{z}$ directions respectively (in all simulations we considered $n_x = n_y$). For a fixed number of points in the $\mathit{z}$ direction, $n_z = 20$, a scan over the following pairs is done in the $\mathit{x}$ and $\mathit{y}$ directions: $(n_x,n_y) = \{(100,100); (200,200); (400,400); (600,600)\}$ until $t=1$. The energy-like quantity $E \equiv \int (\phi^2 + (1 + 1/\tau)^{-1}\, u^2)/2 \, dxdydz$ is conserved by Eqs.~\eqref{DW2_slab}. It is plotted in Fig.~\ref{convergence_energy_nx} as a function of time. Notice that $t=1$ is a Bohm time which is long compared to turbulence time scales which typically range between $1/\rho_\star$ and $\sim 10/\rho_\star $ with $\rho_\star$ being the ratio between the ion larmor radius and the system size. The conservation of $E$ is guaranteed as the grid spacing $\Delta x$ decreases. The line plot on graph~\ref{maxmain_nx} shows that it is effectively the case since in the worst case where $(n_x,n_y) = (100,100)$ the relative change is equal to $4.5 \times 10^{-3}$ and in the well-resolved case where $(n_x,n_y) = (600,600)$ the relative change is almost zero. Accumulating an error of a few percent in a few Bohm times would thus be tolerable when simulating turbulence. 
%\begin{figure}[h!]	
%  \centering
% \hspace*{-0.8cm} \subfloat[]{\includegraphics[width=6.1cm, height=5.8cm]{\RepFigures/FIG22.eps}\label{convergence_energy_nx}}\hspace*{5ex}
% \hspace*{-0.7cm} \subfloat[]{\includegraphics[width=6cm, height=6cm]{\RepFigures/FIG23.eps}\label{maxmain_nx}} 
%  \caption{(a) Conservation of energy with respect to time for different spatial resolutions (b) The relative change in energy with respect to $\Delta x$}
%\end{figure}
For the special case where the box size is $(n_x,n_y,n_z) = (200,200,20)$ shown in the $2$D and $3$D snapshots above, we performed the run up to $t=12$. Though there are fine structures of the grid size scale appearing after $t=6$, the plot in graph~\ref{energy_conservation_relative} shows a relatively good conservation of the energy. The relative variation of energy as a function of time reaches at most $5 \%$.
%\begin{figure}[h!]
%  \centering
% \hspace*{-0.7cm} \includegraphics[width=7cm, height=6cm]{\RepFigures/FIG25.eps}
%  \caption{The relative change in energy with respect to time}
%  \label{energy_conservation_relative} 
%\end{figure}
Similarly, the conservation of the energy is well verified when scanning the number of points in the $\mathit{z}$ direction. A box of size $(n_x,n_y,n_z) = (400,400,n_z)$ is considered where $n_z = \{20,40,60,80,100\}$. In graph~\ref{convergence_energy_nz}, the energy variation converges towards zero as $\Delta z$ decreases. Furthermore, its relative change is in the worst case equal to $5 \times 10^{-4}$ as shown in Fig.~\ref{maxmin}. 
%\begin{figure}[h!]	
%  \centering
% \hspace*{-0.7cm} \subfloat[]{\includegraphics[width=6cm, height=5.6cm]{\RepFigures/FIG26.eps}\label{convergence_energy_nz}}\hspace*{6ex}
% \hspace*{-0.7cm}\subfloat[]{\raisebox{0.4cm}{\includegraphics[width=5.7cm, height=5.35cm]{\RepFigures/FIG27.eps}}\label{maxmin}} 
%  \caption{(a) Plot of the energy as a function of time when scanning the number of points in the $\mathit{z}$ direction (b) The relative change of energy as a function of $\Delta z$}
%\end{figure}

The second target test intends for the demonstration of the convergence of the numerical solution when crossing the separatrix region. For this purpose, a series of simulations were performed with a box of size $(n_x,n_y) = (400,400)$ and a scan over $n_z = \{5,10,20,40,60,80,100\}$. We start by considering the first two solutions given by the simulations having $n_z=5$ and $n_z=10$. The idea is to calculate the difference between the them and repeat the process over each pair of the entire set of solutions $\phi$ given by all the simulations. This is called the \textit{moving difference}. In Fig.~\ref{convergence_nz40}, the moving difference of the solution $\phi$ for $n_z = \{5,10,20,40\}$ at the first poloidal plane $iz = 1$ is plotted in $2$D at time $t=1$. At this stage, for the three pairs given by $\{ \phi(n_z=5),\phi(n_z=10) \}, \{ \phi(n_z=10),\phi(n_z=20) \}, \{ \phi(n_z=20),\phi(n_z=40) \}$, we qualitatively obtain an error that is of the order of the solution. This means that convergence is not yet reached. 
%\begin{figure}[h!]	
%  \centering
%  \includegraphics[width=16cm, height = 17cm]{\RepFigures/FIG28.eps}
%  \caption{Moving difference $\{ \phi(n_z=5),\phi(n_z=10) \}, \{ \phi(n_z=10),\phi(n_z=20) \}, \{ \phi(n_z=20),\phi(n_z=40) \}$: the first two columns show $2$D plots of the solution $\phi$ for two different values of $n_z$. The third column shows a $2$D plot of the difference between the two chosen solutions. Note that in this figure the same color-bar scale is used for all the plots.}
%  \label{convergence_nz40}
%\end{figure}
In Fig.~\ref{convergence_nz100},  the moving difference is calculated for pairs of solutions given by $\{ \phi(n_z=40),\phi(n_z=60) \}, \{ \phi(n_z=60),\phi(n_z=80) \}, \{ \phi(n_z=80),\phi(n_z=100) \}$. We observe that beyond $n_z = 40$ the difference between each pair of solutions is nearly zero which is a good indication that convergence is reached. In other words, for this simulation, one does not need to go beyond $n_z=40$ to study the same physics. This constitutes the main strength of the field-aligned FCI system of coordinates and validates its application to X-point configurations with an exponential decay rate shown in graph~\ref{moving_average}. 
%\begin{figure}[h!]	
%  \centering
% \includegraphics[width=16cm, height = 17cm]{\RepFigures/FIG29.eps}
%  \caption{Moving difference $\{ \phi(n_z=40),\phi(n_z=60) \}, \{ \phi(n_z=60),\phi(n_z=80) \}, \{ \phi(n_z=80),\phi(n_z=100) \}$: the first two columns show $2$D plots of the solution $\phi$ for two different values of $n_z$. The third column shows a $2$D plot of the difference between the two chosen solutions. Note that in this figure the same color-bar scale is used for all the plots.}
%  \label{convergence_nz100}
%\end{figure}
The graph of Fig.~\ref{moving_average} indeed shows a plot line connecting all the fixed averages. It is called the moving average. More specifically, the average is calculated here by dividing the sum of the difference between each pair of solutions by the total number of solutions. It is thus a quantitative mean that allows us to conclude that the numerical solution converges as the resolution in the $\mathit{z}$ direction decreases. In fact, at $n_z = 40$, the norm is approximately $10^{-2}$ which is consistent with the fact that we used second order centered finite differences to compute the parallel gradient operator $\nabla_\parallel$. With this result, we conclude that the numerical solution of the sound-wave propagation problem given an equilibrium with a magnetic island has converged with the expected convergence order. 
%\begin{figure}[h!]	
%  \centering
%  \includegraphics[width=6cm, height = 6cm]{\RepFigures/FIG30.eps}
%  \caption{Average of the norm of the moving difference showing convergence of the numerical solution in $n_z$ at an exponential rate}
%  \label{moving_average}
%\end{figure}
To finish, we do a last test showing the order of convergence of the numerical solution. In fact, for the same set of simulations, i.e: $(n_x,n_y) = (400,400)$ and $n_z = \{5,10,20,40,60,80,100\}$, the test consists of choosing a reference case supposed to be the closest possible case to the analytic solution of the sound-wave problem. With this hypothesis, it is legitimate to calculate the average of the difference between all the simulations and the reference simulation chosen to be that with $n_z=100$ (Root Mean Squared of the difference). The resulting graph~\ref{average_reference_nz100} shows that as $n_z$ tends to $100$, the error tends to $0$ with an estimation of the order of convergence given by the loglog plot in Fig.~\ref{average_reference_nz100}. The convergence is indeed fast, of the order of $a=2.6$, the corresponding slope of the loglog plot. This is once again in good agreement with the use of second order finite differences to compute the parallel gradient operator. To sum up, even with a very elongated island in the perpendicular plane, and a high safety factor, which make the requirements for resolution particularly tough, the latter results show the robustness of the FCI system. At this stage, we can finally conclude that the FCI system of coordinates is quantitatively validated for X-point geometries.
%\begin{figure}[h!]	
%  \centering
%  \subfloat[]{\includegraphics[width=6cm, height=6cm]{\RepFigures/FIG31.eps}\label{average_reference_nz100}}\hspace*{2ex}
%  \subfloat[]{\includegraphics[width=6cm, height=6cm]{\RepFigures/FIG32.eps}\label{average_reference_nz100}} 
%  \caption{(a)Average of the difference between the numerical solution at different $n_z$ values and the solution at the reference case $n_z=100$ (b) loglog plot (natural logarithm) showing the convergence rate having a slope $a=2.6$}
%\end{figure}
%%%%%%%%%%%%%%%%%%%%%%%%%%%%%%%%%%%%%%%%%%%%%%%%%%%%%%%%%%%%%%%%%%%%%%%%
\section{Conclusions\label{conclusions}}
%%%%%%%%%%%%%%%%%%%%%%%%%%%%%%%%%%%%%%%%%%%%%%%%%%%%%%%%%%%%%%%%%%%%%%%%
Many computer resources can be spared by employing coordinate systems which take advantage of the physical characteristics of magnetized plasmas, namely the strong anisotropy of spatial scales between the parallel and the transverse directions (with respect to the equilibrium guiding magnetic field ${\bf B}$). A new three-dimensional Flux-Coordinate Independent system, referred to as FCI in~\cite{Hariri20132419, Hariri2013FENICIA}, is shown to have a number of advantages over earlier approaches. It firstly permits more flexible coding by decoupling the magnetic geometry from the meshgrid. Secondly, it allows for coarse grids since it is field-aligned (one follows the field lines to calculate the parallel gradient operator). Thirdly, the FCI system copes with X-point magnetic geometries, a situation where previous approaches based on magnetic flux coordinates have a fundamental problem due to the singularity of the field-aligned metric. In the present work, the focus is to show that  the FCI approach can be extended to deal with X-point configurations such as magnetic islands. For this purpose, a study of sound-wave propagation across the X-point, in the presence of a magnetic island, was carried out using the code FENICIA (cf. ~\cite{Hariri20132419, Hariri2013FENICIA}). Analytic solutions to the sound-wave problem are constructed both inside and outside the island. The numerical results show the ability of the FCI approach to converge to the derived analytic solution . Then, a test case is considered with an initial perturbation that straddles the magnetic separatrix including the X-point. Since there is no analytic solution in this case, conservation properties were first verified, then three conclusive tests are performed to show the convergence of the numerical solution with respect to the number of points in the $\mathit{z}$ direction (direction of symmetry) allowing one to consider a much thinner grid to study the same Physics. In the latter test, the magnetic island is taken to be very elongated in the perpendicular plane, with a very high safety factor, making the requirements for resolution particularly tough. Even in this very demanding situation, results show the robustness of the FCI approach. We finally conclude this paper by the following: the FCI approach allows not only for a coarser grid by exploiting the anisotropy of the system, but it also allows for flux-coordinate independent operations in the poloidal plane which opens the way to the implementation of complex magnetic geometries free of singularities. It deals without difficulty with X-point configurations and with O-points such as the magnetic axis. The flexible nature of FENICIA presented in~\cite{Hariri20132419, Hariri2013FENICIA} allowed us to demonstrate the application of this coordinate system to a magnetic island in a slab, thus the validity of its application to X-point configurations. The FCI approach can thus be implemented in existing modular codes having pitfalls in dealing with the X-point geometry.
%%%%%%%%%%%%%%%%%%%%%%%%%%%%%%%%%%%%%%%%%%%%%%%%%%%%%%%%%%%%%%%%%%%%%%%%
\section*{Acknowledgments}
%%%%%%%%%%%%%%%%%%%%%%%%%%%%%%%%%%%%%%%%%%%%%%%%%%%%%%%%%%%%%%%%%%%%%%%%
To be added later
%FH would like to thank L. Villard, Y. Peysson, S. Brunner and V. Naulin for providing helpful suggestions concerning the tests performed across the separatrix. FH would also like to express her sincere appreciation to S. Cowley for his valuable feedback on the FCI approach, and his support that benefited this work. Further thanks to X. Litaudon, E. Joffrin, S.-I. Itoh and K. Itoh for their thoughtful advice and very encouraging support.\\
%
%This work, supported by the European Communities under the contract of Association between EURATOM and CEA, was carried out within the framework of the European Fusion Development Agreement. The views and opinions expressed herein do not necessarily reflect those of the European Commission. Part of this work was also supported by the ANR contract E2T2.

%***********************************************************************
% Appendices 
%***********************************************************************
\appendix

%%%%%%%%%%%%%%%%%%%%%%%%%%%%%%%%%%%%%%%%%%%%%%%%%%%%%%%%%%%%%%%%%%%%%%%%
\section{\label{app_A}}
%%%%%%%%%%%%%%%%%%%%%%%%%%%%%%%%%%%%%%%%%%%%%%%%%%%%%%%%%%%%%%%%%%%%%%%%
We seek to define proper coordinates associated to the field lines in the magnetic island equilibrium. For that purpose, let's consider a magnetic field given by Eqs.~\eqref{mag-field} and~\eqref{pol-field}, with 
\begin{equation}
  \psi(x,y) = - \frac{(x-1)^2}{2} + A \cos(y)
\end{equation}
The parallel gradient operators can then be written as:
\begin{equation}
  \nabla_{\parallel} = - [\psi, \cdot ] + \partial_z.
\end{equation}
In order to construct analytic solutions to the above model in the presence of a magnetic island, the idea is to define a coordinate system: $(x,y) \, \longrightarrow \, (\rho',y')$ such that$\rho'$ is a magnetic surface label, i.e: $\nabla_\parallel \rho'= 0$. One would write
\begin{eqnarray}
  \rho' &=& - \psi/A = \frac{(x-1)^2}{2A} - \cos y\\
  y' &=& y 
\end{eqnarray}  
where $\rho'$ is a normalization of $\psi$ so that $\rho' = 1$ refers to the separatrix. Note that in the neighborhood of $y = 0$ (O-point), $\cos y \sim 1- y^2/2 \, \longrightarrow \, \rho' \simeq \frac{(x-1)^2}{2A} -1 + y^2/2$ which defines the equation of an ellipse. With this set of coordinates, the spatial derivatives in the $\mathit{x}$ and $\mathit{y}$ directions write:
\begin{eqnarray}
  \partial_x &=& \partial_x \rho' \, \partial_{\rho'} + \partial_x y' \, \partial_{y'} = x' \, \partial_{\rho'}/A \\
  \partial_y &=& \partial_y \rho' \, \partial_{\rho'} + \partial_y y' \, \partial_{y'} = \sin{y} \, \partial_{\rho'} + \partial_{y'}
\end{eqnarray} 
where $x' = x-1$. This leads to a parallel operator of the form 
\begin{eqnarray}
  \nabla_\parallel &=& x'\, \partial_{y} + A \sin{y}\partial_x + \partial_z\\\nonumber
				  &=& x'(\sin{y} \, \partial_{\rho'} + \partial_{y'}) - x' \sin{y} \, \partial_{\rho'} + \partial_z\\
				  &=& x' \, \partial_{y'} + \partial_z\nonumber
\end{eqnarray}  
The parallel gradient operator can then be written as
\begin{equation}
\nabla_\parallel \phi = x'(\rho', y') \, \partial_{y'} \phi + \partial_z \phi 
\end{equation} 
 where $x'(\rho',y') = \pm \sqrt{(2A)}\, (\rho' + \cos{y'})^{1/2}$. At this point, to ensure the separation of variables, one defines a new system: $(\rho',y') \, \longrightarrow \, (\rho,\eta)$ such that $\rho = \rho'$ and 
\begin{equation}\label{relation_eta}
x'\partial_y \equiv \sqrt{(2A)} \, g(\rho) \, \partial_{\eta} = \pm \sqrt{(2A)} \, (\rho' + \cos{y'})^{1/2} \partial_{y'}
\end{equation}
We conclude from~\eqref{relation_eta} that
\begin{equation}
d\eta = \frac{\pm g(\rho) \, dy'}{(\rho' + \cos{y'})^{1/2}}
\end{equation} 
so, the new variable $\eta$ can be expressed as
\begin{equation}
\eta = \left[ \pm g(\rho) \, \int\limits^y_0 \frac{dy'}{(\rho + \cos{y'})^{1/2}} \right]_{\rho = x'^2/(2A) - \cos{y}} + \eta_0^{\pm}
\end{equation} 
with $g(\rho)$ chosen such that, for all $x'$: 
\begin{eqnarray}
\eta(x',y=\pi) = \pi \\\nonumber
\eta(x',y=-\pi) = -\pi
\end{eqnarray}
This gives 
\begin{equation}
\pi = \pm g\left(1 + \frac{x'^2}{2A}\right) \, \int\limits^{\pi}_0 \frac{dy'}{\left[\cos{y'} + \left( 1+\frac{x'^2}{2A}\right)\right]^{1/2}} \;\; \forall x'
\end{equation} 
so
\begin{equation}
g(\rho)= \pm \pi\, \left[\int\limits^{\pi}_0 \frac{dy'}{\left(\cos{y'} + \rho\right)^{1/2}}\right]^{-1} \;\; \forall x'
\end{equation} 
The function $\eta$ is regular everywhere, but delicate near the X-point where $\rho = 1$ and $\int^y_0 dy'/(\rho' + \cos{y'})^{1/2}$ diverges. When $\rho < 1$, the argument of the integral involves the term $(\cos{y'} + \rho)^{1/2}$ which becomes negative at $\cos{y'} = -\rho$. We choose to limit the integration to the upper bound $y_M$ such that $\cos{y_M} = -\rho$. The situation is described through a schematic of the island in Fig.~\ref{eta_schematic}.
%\begin{figure}[h!]	
%  \centering
%  \includegraphics[width=8cm, height=7cm]{\RepFigures/FIG33.eps}
%  \caption{A schematic of an island with the new variables}
%  \label{eta_schematic}
%\end{figure}
The explicit expression of $g(\rho)$ at the exterior of the island (i.e: $\rho > 1$) can be derived as follows:
\begin{eqnarray}
  \int\limits^{\pi}_0 \frac{dy'}{(\rho + \cos{y'})^{1/2}} &=& \int\limits^{\pi/2}_0 \frac{2d\theta}{(\rho + \cos{2\theta})^{1/2}}\\\nonumber
				  &=& \int\limits^{\pi/2}_0 2d\theta \, \left[\rho +(1-2\,\sin^2{\theta}) \right]^{-1/2} \\
				  &=& \int\limits^{\pi/2}_0 2d\theta \, (\rho + 1)^{-1/2} \, \left[1-2\,\sin^2{\theta}/(1+\rho) \right]^{-1/2} \nonumber
\end{eqnarray} 
Use definitions of elliptic integrals of first kind 
\begin{equation}
K(m) \equiv \int\limits_0^{\pi/2} d\theta \, (1-m\sin^2{\theta})^{-1/2}
\end{equation} 
to get
\begin{equation}
\int\limits_0^\pi dy'\, (\rho + \cos{y'})^{-1/2} = 2 \, (1+\rho )^{1/2} \, K\left(\frac{2}{1+\rho} \right)
\end{equation} 
Note that $K \,(2/(1+\rho))$ diverges logarithmically as $\rho \longrightarrow 1^+$. Finally, for $\rho>1$, $g(\rho)$ can be written as
\begin{equation}
g(\rho) = \frac{\pi}{2} \, (1+\rho )^{1/2} \, \left[ K \, \left( \frac{2}{1+\rho} \right) \right]^{-1}
\end{equation} 

At the interior of the island (i.e: $\rho<1$), however, the limit of integration is $y_M$ such that $\cos{y_M} = -\rho$. Thus
\begin{equation}
\eta' = \pi + \left[ \pm g(\rho) \, \int\limits^y_0 \frac{dy'}{(\rho + \cos{y'})^{1/2}} \right]_{\rho = x'^2/(2A) - \cos{y}}
\end{equation} 
but now 
\begin{equation}
g(\rho)= \frac{\pi}{2} \, \left[\int\limits^{y_M}_0 \frac{dy'}{\left(\cos{y'} + \rho\right)^{1/2}}\right]^{-1} \;\; \forall x'
\end{equation} 
In order to express $g(\rho)$ as a function of $K$, the elliptic integral, we proceed by first rewriting the expression $I = \int^{y_M}_0 \, dy' \,\left(\cos{y'} + \rho\right)^{-1/2}$ by change of variables. We indeed have
\begin{equation}
\cos{y'} = 1 -(1+\rho) \sin^2{\alpha}
\end{equation} 
where $0\leq \alpha \leq \pi/2$. Thus
\begin{equation}
(\cos{y'}  + \rho)^{-1/2}= (1+\rho)^{-1/2} (1-\sin^2{\alpha})^{-1} = (1+\rho)^{-1/2} (\cos{\alpha})^{-1}
\end{equation}  
Moreover, by differentiation we get
\begin{equation}
-\sin{y'} \, dy' = -2 (1+\rho) \sin{\alpha}\cos{\alpha}\, d\alpha
\end{equation} 
Also,
\begin{equation}
\sin^2{y'} = 1-\cos^2{y'} = 2\,(1 +\rho) \, \sin^2{\alpha} \, \left[ 1 - \frac{1+\rho}{2} \, \sin^2{\alpha} \right]
\end{equation} 
yielding
\begin{equation}
\sin{y'} = \left[ 2\,(1 +\rho)\right]^{1/2} \, \sin{\alpha} \, \left[ 1 - \frac{1+\rho}{2} \, \sin^2{\alpha} \right]^{1/2}
\end{equation} 
and 
\begin{equation}
dy' = \left[ 2\,(1 +\rho)\right]^{1/2} \, \cos{\alpha} \, \left[ 1 - \frac{1+\rho}{2} \, \sin^2{\alpha} \right]^{-1/2}\, d\alpha
\end{equation} 
Hence, 
\begin{eqnarray}
I &=& \sqrt{2} \, \int\limits^{\pi/2}_0 \frac{d\alpha}{\left[1 -\frac{1+\rho}{2} \sin^2{\alpha}\right]^{1/2}}\\
&=& \sqrt{2} \,K \left(\frac{1+\rho}{2}\right)
\end{eqnarray} 
Finally for $\rho<1$ we have,
\begin{equation}
g(\rho) = \frac{\pi}{2^{3/2}} \,\left[ K \, \left( \frac{1 + \rho}{2} \right) \right]^{-1}
\end{equation} 
For $x'<0$, as shown in Fig.~\ref{eta_schematic}, $\eta'$ needs to be completed. We define
\begin{equation}
\eta' = \pi - \left[ g(\rho) \, \int\limits^y_0 \frac{dy'}{(\rho + \cos{y'})^{1/2}} \right]_{\rho = x'^2/(2A) - \cos{y}}
\end{equation} 
so that $\pi/2 \leq \eta' \leq 3\pi/2$ when $x'<0$. And
\begin{equation}
\eta' = 2\pi + \left[ g(\rho) \, \int\limits^y_0 \frac{dy'}{(\rho + \cos{y'})^{1/2}} \right]_{\rho = x'^2/(2A) - \cos{y}}
\end{equation} 
for $x'>0$, but $y<0$ if one wants  $3\pi/2 \leq \eta' \leq 2\pi$ in the fourth quadrant.

%***********************************************************************
%% References
%***********************************************************************

%% Following citation commands can be used in the body text:
%% Usage of \cite is as follows:
%%   \cite{key}         ==>>  [#]
%%   \cite[chap. 2]{key} ==>> [#, chap. 2]
%% References with bibTeX database:
%\bibliographystyle{elsarticle-num} % or unsrt
%\bibliography{bibmain}

% References POP

\end{document}